# Virtual staining of defocused autofluorescence images of unlabeled tissue using deep neural networks

Yijie Zhang[1,2,3†], Luzhe Huang[1,2,3†], Tairan Liu[1,2,3], Keyi Cheng[4], Kevin de Haan[1,2,3], Yuzhu Li[1,2,3], Bijie Bai[1,2,3], and Aydogan Ozcan[1,2,3,5,*]

[1]Electrical and Computer Engineering Department, University of California, Los Angeles, CA 90095, USA

[2]Department of Bioengineering, University of California, Los Angeles, CA 90095, USA

[3]California NanoSystems Institute, University of California, Los Angeles, CA 90095, USA

[4]Department of Mathematics, University of California, Los Angeles, CA 90095, USA

[5]Department of Surgery, David Geffen School of Medicine, University of California, Los Angeles, CA 90095, USA

*Correspondence: ozcan@ucla.edu

[†]Equally contributing authors

# Abstract

Deep learning-based virtual staining was developed to introduce image contrast to label-free tissue sections, digitally matching the histological staining, which is time-consuming, labor-intensive, and destructive to tissue. Standard virtual staining requires high autofocusing precision during the whole slide imaging of label-free tissue, which consumes a significant portion of the total imaging time and can lead to tissue photodamage. Here, we introduce a fast virtual staining framework that can stain defocused autofluorescence images of unlabeled tissue, achieving equivalent performance to virtual staining of in-focus label-free images, also saving significant imaging time by lowering the microscope's autofocusing precision. This framework incorporates a virtual-autofocusing neural network to digitally refocus the defocused images and then transforms the refocused images into virtually stained images using a successive network. These cascaded networks form a collaborative inference scheme: the virtual staining model regularizes the virtual-autofocusing network through a style loss during the training. To demonstrate the efficacy of this framework, we trained and blindly tested these networks using human lung tissue. Using 4× fewer focus points with 2× lower focusing precision, we successfully transformed the coarsely-focused autofluorescence images into high-quality virtually stained H&E images, matching the standard virtual staining framework that used finely-focused autofluorescence input images. Without sacrificing the staining quality, this framework decreases the total image acquisition time needed for virtual staining of a label-free whole-slide image (WSI) by ~32%, together with a ~89% decrease in the autofocusing time, and has the potential to eliminate the laborious and costly histochemical staining process in pathology.

# Introduction

Histological analysis is considered to be the gold standard for tissue-based diagnostics. In the histological staining process, the tissue specimen is first sliced into 2–10 μm thin sections and then fixed on microscopy slides. These slides are stained in a process that dyes the specimen with markers by binding e.g., chromophores to different tissue constituents, revealing the sample's cellular and subcellular morphological information under a microscope [1]. However, the traditional histological staining is a costly and time-consuming procedure. Some types of stains, such as immunohistochemical staining (IHC), require a specialized laboratory infrastructure and skilled histotechnologists to perform tissue preparation steps.

The ability to virtually stain microscopic images of unlabeled tissue sections was demonstrated through deep neural networks, avoiding the laborious and time-consuming histochemical staining processes. These deep learning-based label-free virtual staining methods can use different input imaging modalities, such as autofluorescence microscopy [2–4], hyperspectral imaging [5], quantitative phase imaging (QPI) [6], reflectance confocal microscopy [7], and photoacoustic microscopy [8], among others [9–11]. Virtual staining, in general, has the potential to be used as a substitute for histochemical staining, providing savings in both costs and tissue processing time. It also enables the preservation of tissue sections for further analysis by avoiding destructive biochemical reactions during the chemical staining process [12].

In all these label-free virtual staining methods, the acquisition of in-focus images of the unlabeled tissue sections is essential. In general, focusing is a critical but time-consuming step in scanning optical microscopy used to correct focus drifts caused by mechanical or thermal fluctuations of the microscope body and the nonuniformity of the specimen's topology [13]. Focus map surveying, the most adopted autofocusing method for whole slide imaging of tissue sections, creates a pre-scan focus map by sampling the focus points in a pattern [14]. At each focus point of this pattern, the autofocusing process captures an axial stack of images, from which it extracts image sharpness measures at different axial depths and locates the best focal plane using an iterative search algorithm [15–17]. To acquire finely-focused whole slide images (WSI) of label-free tissue sections and generate high-quality virtually stained images, standard virtual staining methods demand many focus points across the whole slide area with high focusing precision to form an accurate pre-scan focus map. However, this fine focus search process is time-consuming to perform across a WSI and might introduce photodamage and photobleaching [18] on the tissue sample. To alleviate some of these problems, recent works in optical microscopy have explored the use of deep learning for online autofocusing [19–22], offline autofocusing [23], and depth-of-field (DoF) enhancement [24–26]. Despite all this progress, integrating deep learning-based autofocusing methods with virtual staining of unstained tissue remains to be explored.

Here, we demonstrate a deep learning-based fast virtual staining framework that can generate high-quality virtually stained images using defocused autofluorescence images of label-free tissue. As shown in Fig. 1, this framework uses an autofocusing neural network (Deep-R) [23] to digitally refocus the defocused autofluorescence images. Then a virtual staining network is used to transform the refocused images into virtually stained images, matching the brightfield microscopic images of the histochemically stained tissue (ground truth). Instead of training the two cascaded networks (i.e., the autofocusing and virtual staining neural networks) separately, we first trained the virtual staining network and used the learned virtual staining model to regularize the Deep-R network using a style loss during the training stage, which formed a collaborative inference scheme.

To demonstrate the success of this deep learning-based fast virtual staining framework, we trained the networks using human lung tissue sections. Through blind testing on coarsely-focused autofluorescence images of unlabeled lung tissue sections, the fast virtual staining framework successfully generated virtual H&E stained images matching the staining quality of the standard virtual staining framework that used in-focus autofluorescence images of the same samples. These coarsely-focused autofluorescence images of unlabeled tissue were acquired with 4× fewer focus points and 2× lower focusing precision than their finely-focused counterparts used in the standard virtual staining framework; this resulted in a ~32% decrease in the total image acquisition time (per WSI) and a ~89% decrease in autofocusing time using a benchtop scanning optical microscope. With its capability to stain defocused images of unstained tissue, we believe this virtual staining method will save time without sacrificing the image quality of the virtually stained images and be highly useful for histology.

## Results

The standard virtual staining framework [2] uses in-focus autofluorescence microscopic images of label-free tissue to digitally stain the corresponding images. To generate high-quality virtually stained images using *defocused* autofluorescence images, we first use a Deep-R [23] network for virtual autofocusing, followed by the virtual staining of the resulting refocused autofluorescence images, as shown in Fig. 1 and Fig. 2(a). To achieve accurate virtual staining on the refocused autofluorescence images (the output of Deep-R), the trained virtual staining model was used to regularize the Deep-R network during its training by introducing a style loss, which minimizes the difference between multiscale virtual staining features of the Deep-R output and the target (see the Materials and Methods section for details). In other words, the presented defocused image virtual staining framework does *not* involve a simple cascade of two different, separately trained neural networks, one following another.

We trained this defocused image virtual staining framework with a dataset of 5832 human lung tissue fields-of-views (FOVs), each of which had 512×512 pixels, imaged using a 40×/0.95 NA objective lens. As shown in Fig. 1(b), to train the Deep-R network, it was fed with accurately paired image data consisting of (1) autofluorescence images of label-free tissue (including DAPI and TxRed filter channels) acquired at different axial defocus distances (ranging from -2 µm to 2 µm with an axial step size of 0.5 µm, as illustrated in Fig. 1(b)) as inputs, and (2) the corresponding in-focus DAPI and TxRed autofluorescence images as targets. During the training of the Deep-R network, the input autofluorescence images (defocused) in each batch were randomly picked from the z-stacks. To train the virtual staining network, registered pairs of in-focus autofluorescence images (DAPI and TxRed channels) captured before the histochemical staining and the brightfield images of the same tissue sections after their histochemical staining are used (see Fig. 1(a) and the Materials and Methods section). The two networks (Deep-R and the successive virtual staining network) are linked together by a style loss during the training (see the Materials and Methods section), forming a collaborative inference scheme.

Once trained, the defocused image staining framework can generate high-quality virtually stained images using defocused autofluorescence microscopic images of label-free tissue as its input; this capability enables using fewer autofocus points and lower focusing precision at each focus point during the WSI scanning process. To demonstrate its success, we blindly tested and compared the performance of the standard in-focus image virtual staining framework and our defocused image virtual staining framework on 2081 unique image FOVs (each image with 2048 x 2048 pixels) from ten new patients that were never seen by the network before. For the standard in-focus virtual staining framework, we acquired finely-focused whole slide autofluorescence images of the test tissue sections (~23 mm$^2$ of sample area per patient on average) by using focus points at 8.5% of the total acquired image FOVs, and a ±0.35 µm focusing precision at each focus point to form a fine focus map before the WSI scanning. On the other hand, for the defocused virtual staining framework, we used a smaller number of focus points that only took up 2.1% of the total acquired image FOVs, and reduced the focusing precision to ±0.83 µm for each focus point to acquire coarsely-focused whole slide autofluorescence images, as illustrated in Fig. 2(a). These changes reduced the autofocusing time (per WSI) from 9.8 minutes to 1.1 minutes and the total image acquisition time from 27.1 minutes to 18.4 minutes, achieving an 88.8% decrease in the autofocusing time and a 32.1% decrease in the entire image acquisition process per WSI (see the Materials and Methods section). Because of the coarse focus map, the acquired autofluorescence image FOVs exhibit various defocus distances for each WSI. Figure 2(b) presents the zoomed-in regions of the acquired autofluorescence images, and the generated virtually stained images. Both frameworks (in-focus vs. defocused image virtual staining networks) can generate high-quality staining that presents a good match to the corresponding histochemically stained ground truth images. Although the fast virtual staining framework took defocused autofluorescence images as its input, with an apparent loss of sharpness and contrast compared to their finely-focused counterparts, it can still achieve comparable virtual staining performance to the standard network that used in-focus input images.

To further showcase the ability of the presented framework, we compared the performance of the standard in-focus image virtual staining network (termed framework 1) and the fast, defocused image virtual staining network (termed framework 2) using the same coarsely-focused autofluorescence images as input. Framework 1 directly applies the virtual staining network on these coarsely-focused autofluorescence images of label-free tissue sections, without using the Deep-R network for refocusing, whereas framework 2 uses the Deep-R network for refocusing of defocused autofluorescence images, followed by the virtual staining. In this comparison, the results of the standard virtual staining using finely-focused image FOVs were also used as a baseline, which we termed framework 3. Figure 3 reports a detailed comparison of these three frameworks' inference on various FOVs of different lung tissue sections, never used during the training phase. Using defocused autofluorescence images, framework 1 presented a noticeable sharpness and contrast degradation in its virtually stained images (Fig. 3(a-e)). Furthermore, it also caused hallucinations of nuclei and red blood cells, and related artifacts that cannot be seen in the results of framework 3 (Fig. 3(k-o)). In contrast, framework 2 (Fig. 3(f-j)) successfully avoided these hallucinations and artifacts, and produced sharp virtually stained images that are in-focus, with a good match to the results of framework 3, further confirming the conclusions reported in Fig. 2(b).

We further quantified the virtual staining performance by calculating the peak signal-to-noise ratio (PSNR) and the structural similarity index (SSIM) [27] between (1) framework 1 or framework 2 and (2) the corresponding baseline results in framework 3, as shown in Fig. 3(a-j). Compared to the inference of framework 1, both metrics (PSNR and SSIM) were significantly improved by the reported defocused image virtual staining method (framework 2), demonstrating its robustness to defocused image inputs.

Since the chromatic contrast among different tissue components serves as one of the most significant features/cues for pathologists to interpret tissue sections, we also quantified the color distribution of the virtually stained images by converting them from RGB to YCbCr color space and then plotting the histograms of Cb and Cr channels, as shown in Fig. 3(p-t). The two chroma components (Cb and Cr) can present the blue and red information of the virtually stained images, respectively, reflecting the staining quality of H&E where hematoxylin stains nuclei a purplish-blue, and eosin stains the extracellular matrix and cytoplasm pink [28]. For both Cb and Cr channels, the distributions of framework 1 image inference have obvious shifts compared to the other two frameworks (2 and 3). In contrast, the image inference results of framework 2 agree well with the distributions of framework 3, further validating the success of our defocused image virtual staining framework. It is also worth noting that the defocused image virtual staining has performance degradation on autofluorescence images with a large defocus distance (e.g., with an axial defocus amount of > 2.5 µm, see Fig. 3(i-j)), which is not surprising since this large defocus lies outside of its training range (± 2 µm).

Next, we used the differences in the YCbCr color space to further quantify the relationship between the virtual staining performance and the axial image defocus distance; see Fig. 4. To conduct this analysis, we acquired z-stacks of autofluorescence images of label-free lung tissue sections (over an axial range of -3 µm to 3 µm with a step size of 0.5 µm, see the Materials and Methods section), resulting in 562 unique image FOVs, each with 512×512 pixels. Then we separately tested framework 1 and framework 2 on the acquired autofluorescence images at different axial defocus distances. The resulting virtually stained images were used to compute the absolute YCbCr color difference with respect to the virtual staining results of framework 3 that used in-focus autofluorescence images (i.e., using z = 0 µm of the same FOVs). The average absolute color differences (of the 562 image FOVs) were plotted as a function of the axial defocus distance, as shown in Fig. 4(a). These results reveal that framework 2 performs similarly to framework 1 when the input autofluorescence image has a small defocus distance (e.g., < 1 µm). On the other hand, when the input autofluorescence images have a large defocus distance of ≥ 1 µm, framework 2 has significantly better performance than framework 1. We also performed paired upper-tailed $t$ tests (see the Materials and Methods section) between the two frameworks for each color channel and plotted the resulting $p$ values as a function of the axial defocus distance to further illustrate the performance improvement of our defocused image virtual staining framework, as presented in Fig. 4(b). The $t$ test results demonstrate that the defocused image virtual staining network has a statistically significant improvement in virtual staining performance over the standard virtual staining network at a defocus distance of ± ~1 µm or larger.

## Discussion

We demonstrated a deep learning-based framework that decreases the total image acquisition time needed for virtual staining of a label-free WSI by ~32%, also resulting in a ~89% decrease in the autofocusing time per tissue slide. By combing a Deep-R network with a virtual staining model, our framework generated virtual H&E stained images from coarsely-focused whole slide autofluorescence images of label-free tissue sections, matching the standard virtual staining inference that used finely-focused WSIs that were acquired with 4× more focus points and 2× higher focusing precision. For a 1 cm$^2$ label-free tissue section, consisting of 900 image FOVs (with each FOV having 2048×2048 pixels), the data acquisition for in-focus autofluorescence images per sample takes ~6900 seconds using a scanning optical microscope (see the Materials and Methods section). With the help of the presented defocused image virtual staining network, this scanning time can be reduced to ~4800 seconds by acquiring coarsely-focused WSIs with reduced focus points and focusing precision. After the training is complete, which is a one-time effort, the inference process for both the Deep-R and the virtual staining network only takes ~18 seconds (i.e., ~36 seconds in total) for a tissue area of 1 cm$^2$; stated differently, the total inference time for virtual staining is negligible compared with the whole slide image acquisition process.

Besides saving significant amounts of image acquisition time, the framework presented here can also act as an add-on module to improve the robustness of the standard virtual staining framework. Even when using high-precision pre-scan focus maps, parts of the WSI can still be inaccurately focused due to fluctuations of the microscope body and local variations of the specimen's topology. This can cause either defocused image FOVs in parts of the WSI or an inaccurate focus map. Our fast virtual staining framework can be applied to these defocused FOVs to generate the same high-quality virtually stained images as the other in-focus regions, improving the inference consistency of the virtual staining framework using label-free tissue sections.

The ability of the fast virtual staining framework to generate high-quality stained images using coarsely-focused autofluorescence images stems from the integration of the Deep-R and virtual staining neural networks. In the training of these two networks, the style loss (see Fig. 5(a) and the Materials and Methods section) serves as an essential regularization term to optimize the Deep-R network to generate refocused autofluorescence images suitable for the trained virtual staining network. Since the neural network output usually has a distribution deviation from its target, directly feeding the refocused output images of Deep-R into the virtual staining network leads to artifacts and hallucinations in the generated virtually stained images. The style loss, however, helps us regularize the Deep-R network such that Deep-R learns to recover the "style features" used for the virtual staining network, close to those of the in-focus autofluorescence images. Furthermore, conventional loss terms such as the mean absolute error (MAE), etc., emphasize low-level image features by achieving pixel-wise correlations between the output and the ground truth images. The style loss, however, penalizes the high-level image features by comparing multiscale features of the virtual staining network, enabling Deep-R to retrieve the features for the connected virtual staining network.

By introducing an additional Deep-R network in the inference process, the fast, defocused image virtual staining framework can be implemented on conventional fluorescence microscopes without hardware modifications or a customized optical setup. This fast virtual staining workflow can also be expanded to many other stains, such as Masson's Trichrome stain, Jones' silver stain, and immunohistochemical (IHC) stains [2–4, 12]. Although the virtual staining approach presented here was demonstrated based on the autofluorescence imaging of unlabeled tissue sections, it can also be used to speed up the virtual staining workflow of other label-free microscopy modalities [6, 7].

## Materials and Methods

### *Image data acquisition*

The neural networks were trained using microscopic images of thin tissue sections from lung needle core biopsies. Unlabeled tissue sections were obtained from existing deidentified specimens from the UCLA Translational Pathology Core Laboratory (TPCL). The human lung tissue blocks were sectioned using a microtome into ~4 μm thick sections, then deparaffinized using xylene and mounted on a standard glass slide using mounting medium Cytoseal 60 (Thermo-Fisher Scientific). The autofluorescence images were captured using a Leica DMI8 microscope, controlled with Leica LAS X microscopy automation software. The unstained tissue sections were excited near the ultraviolet range and imaged using a DAPI filter cube (Semrock OSFI3-DAPI5060C, EX377/50 nm EM 447/60 nm) as well as a TxRed filter cube (Semrock OSFI3-TXRED-4040C, EX 562/40 nm EM 624/40 nm). The autofluorescence images were acquired with a 40×/0.95 NA objective (Leica HC PL APO 40×/0.95 DRY). Each FOV was captured with a scientific complementary metal-oxide-semiconductor (sCMOS) image sensor (Leica DFC 9000 GTC) with an exposure time of ∼100 ms for the DAPI channel and ∼300 ms for the TxRed channel.

While acquiring the autofluorescence images of the samples used to train the networks, we first built a fine pre-scan focus map with focus points uniformly distributed over the sample, taking up ~10% of the total image FOVs. Each focus point had a focusing precision of ±0.35 μm. At each FOV, we acquired a z-stack of autofluorescence images ranging from −2 to 2 μm with 0.5 μm axial spacing, where z = 0 μm refers to the in-focus position from the fine pre-scan focus map. The in-focus autofluorescence images (z = 0 μm) are used as the network input to the virtual staining network (Fig. 1(a)) and network target for the Deep-R network (Fig. 1(b)). The autofluorescence images at different axial depths in the z-stack were randomly fed into the Deep-R network as input.

For each testing tissue sample, we first built the fine pre-scan focus map similar to the acquisition of the training images, and acquired the finely-focused whole slide autofluorescence images, which were the test inputs for the standard virtual staining framework. Then we built a coarse pre-scan focus map and acquired the corresponding coarsely-focused WSI for the same sample. The focus points on the coarse focus map had a precision of ±0.83 μm and were evenly distributed over the sample, taking up ~2% of total image FOVs. For the blind testing samples that were used to quantify the relationship between the virtual staining performance and the axial defocus distance in Fig. 4, we built fine pre-scan focus maps, the same as the acquisition of the training sample. At each image FOV, we then acquired z-stacks ranging from −3 to 3 μm with 0.5 μm axial spacing. To achieve the ±0.35 μm (or ±0.83 μm) focusing precision, the Leica LAS X microscopy automation software performs a two-step search algorithm to find the in-focus position. It first controls the microscope to implement a coarse focus search in a z-stack that has a range of 50 μm with 23 (or 9) axial steps. Then a fine focus search in a z-stack with a range of 20 μm with 29 (or 13) axial steps finds the optimal focus; the time of the autofocusing process at each focus point is 33 (or 15) seconds, respectively.

After the autofluorescence imaging of each tissue section, the H&E histochemical staining was performed by UCLA TPCL. These stained slides were then digitally scanned using a brightfield scanning microscope (Leica Biosystems Aperio AT2), which were used as ground truth images.

*Image pre-processing and co-registration*

To train the network through supervised learning, matching pairs of images must be obtained before and after the histochemical staining. To do this, the in-focus autofluorescence images of an unlabeled tissue section were co-registered to brightfield images of the same tissue section after it was histochemically stained. This image co-registration was done through a combination of coarse and fine matching steps that were used to progressively improve the alignment until sub-pixel level accuracy was achieved, which followed the process reported by Rivenson et al. [2]. In the coarse image registration, a cross-correlation-based method was first used to extract the most similar portions in the stained images matching the autofluorescence images. Next, multimodal image registration [29] between the extracted histochemically stained images and the autofluorescence images resulted in an affine transformation, which was applied to the extracted stained images to correct any changes in size or rotation. To achieve pixel-level co-registration accuracy, a fine matching step using an elastic pyramidal registration algorithm [30, 31] was implemented. Since this step relies upon local-correlation-based matching, an initial rough virtual staining network is applied to the autofluorescence images. These roughly stained images were then co-registered to the brightfield images of the histochemically stained tissue using the elastic pyramidal registration algorithm.

Before feeding the aligned images into the neural networks, several pre-processing steps were applied to the images. For the Deep-R network, each pair of input and target autofluorescence images were normalized to have zero mean and unit variance. The same normalization was also applied to the input autofluorescence images of the virtual staining network. The histochemically stained images (ground truth) were converted to the YCbCr color space before being fed into the virtual staining network as target. For both the Deep-R and virtual staining networks, all image pairs were randomly partitioned into patches of 512×512 pixels and then augmented eight times by random flipping and rotations during training.

*Network architecture, training, and validation*

To perform the virtual staining network, we used a GAN [32] architecture (see Fig. 5(b)), which is composed of two deep neural networks, including a generator and a discriminator. The generator network follows a U-net [33] structure, consisting of four downsampling blocks with residual connections and four upsampling blocks. Each downsampling block comprises three convolution layers and their activation functions, which double the number of channels. An average pooling layer follows these convolution layers with a stride and kernel size of two. The upsampling blocks first 2× bilinearly resize the tensors and then use three convolution layers with activation functions to reduce the number of channels by a factor of four. Skip connections between the downsampling and the upsampling layers at the same level allow features at various scales to be learned.

The input of the discriminator network was either the virtually stained images from the generator or the histochemically stained ground truth images. The discriminator contains six convolution blocks, each of which consists of two convolution layers that double the number of channels and has a stride of two. These six blocks were followed by a global pooling layer and two dense layers to generate a scalar after a sigmoid activation function.

During the training phase, the virtual staining network iteratively minimizes the loss functions of the generator and discriminator networks, defined as:

$$L_{G_{VS}} = L_{MAE}(z, G_{VS}(y)) + \eta L_{adv}(G_{VS}(y)) + \lambda L_{TV}(G_{VS}(y)) \quad (1)$$

$$L_{D_{VS}} = D_{VS}(G_{VS}(y))^2 + (1 - D_{VS}(z))^2 \quad (2)$$

$D_{VS}(.)$ and $G_{VS}(.)$ refer to the outputs of the discriminator and generator for the virtual staining network, respectively. $y$ represents the in-focus autofluorescence images, and $z$ denotes the brightfield counterparts of the histochemically stained tissue (ground truth). In these loss functions, the total variation (TV) and MAE loss terms are used as structural regularization terms to ensure that highly accurate virtually stained images are generated. The MAE loss and TV operator are defined as:

$$L_{MAE}(z, G_{VS}(y)) = \frac{1}{P \times Q} \sum_p \sum_q |z_{p,q} - G_{VS}(y)_{p,q}| \quad (3)$$

$$L_{TV}(G_{VS}(y)) = \sum_p \sum_q (|G_{VS}(y)_{p+1,q} - G_{VS}(y)_{p,q}| + |G_{VS}(y)_{p,q+1} - G_{VS}(y)_{p,q}|) \quad (4)$$

P and Q represent the number of vertical and horizontal pixels of the image patch, and $p$ and $q$ represent the pixel locations. The adversarial loss is defined as:

$$L_{adv}(G_{VS}(y)) = (1 - D_{VS}(G_{VS}(y)))^2 \quad (5)$$

The regularization parameters ($\eta$ and $\lambda$) were empirically set to 2000 and 0.02.

For the virtual staining network, the generator and discriminator both use the Adam [34] optimizer with the initial learning rates of $10^{-4}$ and $10^{-5}$, respectively. A batch size of 4 was used during the training phase, and the training process converged after ~40,000 iterations (equivalent to ~30 epochs).

For the Deep-R network, a similar GAN structure to the virtual staining network was used, but several modifications were made to the generator architecture. The Deep-R generator adapts five upsampling and downsampling blocks, and each downsampling or upsampling block contains two convolution layers in conjunction with a residual connection. In the downsampling path, instead of an average pooling layer, the Deep-R generator adapts max-pooling layers. For the objective function of the generator training, we used an adversarial loss from the discriminator, a perceptual loss [35], and a style loss based on high-level image features, in addition to the MAE loss and the multiscale structural similarity (MSSSIM) losses between the Deep-R output and the ground truth in-focus images, as shown in Fig. 5(a). The generator and discriminator losses of the Deep-R network are defined as:

$$L_{G_{DR}} = aL_{adv}(G_{DR}(x)) + bL_p(y, G_{DR}(x)) + cL_s(y, G_{DR}(x))$$
$$+ dL_{MAE}(y, G_{DR}(x)) + eL_{MSSSIM}(y, G_{DR}(x)) \quad (6)$$

$$L_{D_{DR}} = D_{DR}(G_{DR}(x))^2 + (1 - D_{DR}(y))^2 \quad (7)$$

*a, b, c, d, e* are training coefficients empirically set as 300, 2000, 500, 100, 100, respectively. $D_{DR}(.)$ and $G_{DR}(.)$ refer to the discriminator and generator outputs for the Deep-R, respectively. *x* denotes the autofluorescence images taken from a z-stack ranging from −2 to 2 μm with an axial step size of 0.5 μm; same as before, *y* represents the in-focus autofluorescence images. The adversarial loss and MAE loss are defined as before. The perceptual loss $L_p$ is defined as

$$L_p(y, G_{DR}(x)) = \frac{1}{K} \sum_k |D_{DR,k}(y) - D_{DR,k}(G_{DR}(x))|_1 \quad (8)$$

Where $D_{DR,k}(\cdot)$ represents the output feature map at the *k*-th convolutional block of the discriminator. The style loss $L_s$ is defined as

$$L_s(y, G_{DR}(x)) = \frac{1}{M} \sum_{m=1}^{M} |G_{VS,m}(y) - G_{VS,m}(G_{DR}(x))|_1 \quad (9)$$

$G_{VS,m}(\cdot)$ stands for the output feature map at the *m*-th downsampling block of the trained virtual staining network (see Fig. 5(a)).

The MSSSIM loss, $L_{MSSSIM}$, is defined as:

$$L_{MSSSIM}(f,g) = 1 - \left[\frac{2\mu_{f_S}\mu_{g_S} + C_1}{\mu_{f_S}^2 + \mu_{g_S}^2 + C_1}\right]^{\alpha_S} \times \Pi_{j=1}^{S}\left[\frac{2\sigma_{f_j}\sigma_{g_j} + C_2}{\sigma_{f_j}^2 + \sigma_{g_j}^2 + C_2}\right]^{\beta_j}\left[\frac{\sigma_{f_jg_j} + C_3}{\sigma_{f_j}\sigma_{g_j} + C_3}\right]^{\gamma_j} \quad (10)$$

where $f_j$ and $g_j$ are the distorted (or recovered/inferred) and reference images downsampled $2^{j-1}$ times, respectively; $\mu_f, \mu_g$ are the averages of $f, g$; $\sigma_f^2, \sigma_g^2$ are the variances of $f, g$, respectively; $\sigma_{fg}$ is the covariance of $f, g$; $C_1, C_2, C_3$ are the constants used to stabilize the division with a small denominator; and $\alpha_S, \beta_j, \gamma_j$ are exponents used to adjust the relative importance/weights of different components. The MSSSIM function is implemented using the TensorFlow function tf.image.ssim_multiscale using its default parameter settings.

The generator and discriminator for the Deep-R network used the Adam optimizers with the initial learning rates of $10^{-5}$ and $10^{-6}$, respectively. We used a batch size of 5 in our training phase, and the training process converged after ~100,000 iterations (equivalent to ~10 epochs).

*Quantitative image metrics*

PSNR is defined as:

$$\text{PSNR} = 10 \times \left(\frac{\text{MAX}_I^2}{\text{MSE}}\right) \quad (11)$$

where $\text{MAX}_I$ is the maximum possible value of the ground truth image. MSE is the mean square error between the two images being compared, defined as:

$$\text{MSE} = \frac{1}{n^2}\sum_{i=0}^{n-1}\sum_{j=0}^{n-1}[\text{I}(i,j) - \text{K}(i,j)]^2 \quad (12)$$

where I is the target image, and K is the image that is compared with the target image.

SSIM is defined as:

$$\text{SSIM}(a,b) = \frac{(2\mu_a\mu_b + C_1)(2\sigma_{a,b} + C_2)}{(\mu_a^2 + \mu_b^2 + C_1)(\sigma_a^2 + \sigma_b^2 + C_2)} \tag{13}$$

where $a$ and $b$ are the two images being compared. $\mu_a$ and $\mu_b$ are the mean values of $a$ and $b$, respectively. $\sigma_a$ and $\sigma_b$ are the standard deviations of $a$ and $b$, respectively. $\sigma_{a,b}$ is cross-covariance of $a$ and $b$. $C_1$ and $C_2$ are the constants that are used to avoid division by zero.

*Statistical analysis*

Paired upper-tailed *t* tests were used to determine whether statistically significant improvements were made when using the fast, defocused virtual staining framework. For each YCbCr color channel and the axial defocus distance, the paired upper-tailed *t* test was performed across the 562 unique FOVs using the absolute color differences between the virtual staining results of framework 1 and framework 3 (termed c1: comparison 1), and the absolute color differences between the virtual staining results of framework 2 and framework 3 (termed c2: comparison 2). The null hypothesis for the paired upper-tailed *t* test is that c1 and c2 have the same mean. We used a 0.05 statistical significance to reject the null hypothesis in favor of an alternate upper-tailed hypothesis that c2 has a smaller mean than c1, indicating that the framework 2 has a statistically significant improvement over the framework 1 (see Fig. 4).

*Implementation details*

The image pre-processing was implemented in MATLAB using version R2018b (MathWorks). The neural networks were implemented using Python version 3.9.0 and TensorFlow 2.1.0. The training was performed on a desktop computer with an Intel Xeon W-2265 central processing unit (CPU), 256 GB random-access memory (RAM), and an Nvidia GeForce RTX 2080 TI graphics processing unit (GPU).

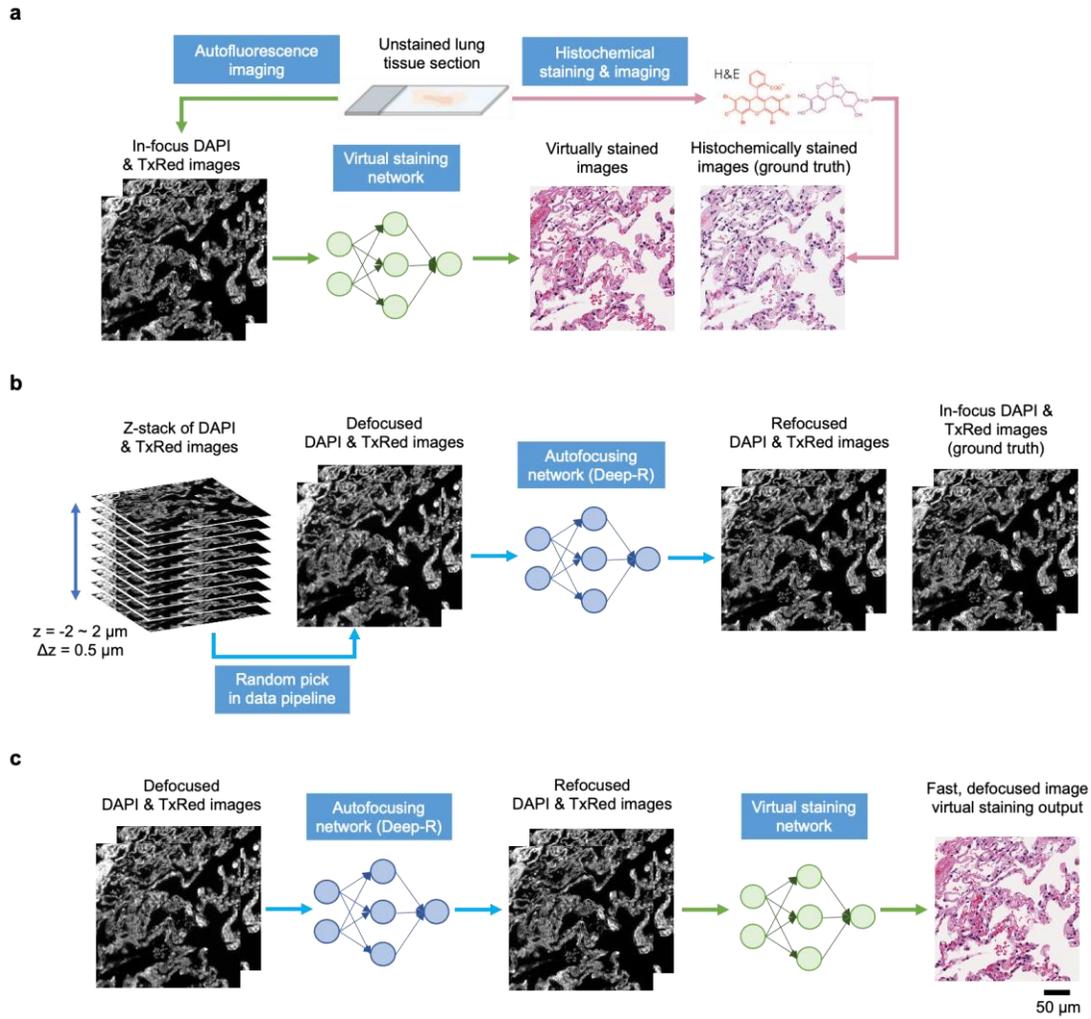

**Figure 1 The diagram of the training and testing schemes for the defocused image virtual staining framework**. **a**. Training of the standard in-focus image virtual staining network. Aligned pairs of in-focus autofluorescence images captured before the histochemical staining process and the brightfield images of the same tissue sections after the histochemical staining are used. **b**. Training of the autofocusing network (Deep-R). Autofluorescence images (defocused) were randomly picked from z-stacks (ranging from -2 μm to 2 μm) as the network inputs. The network target is the corresponding in-focus autofluorescence image. **c**. Testing of the defocused image virtual staining framework.

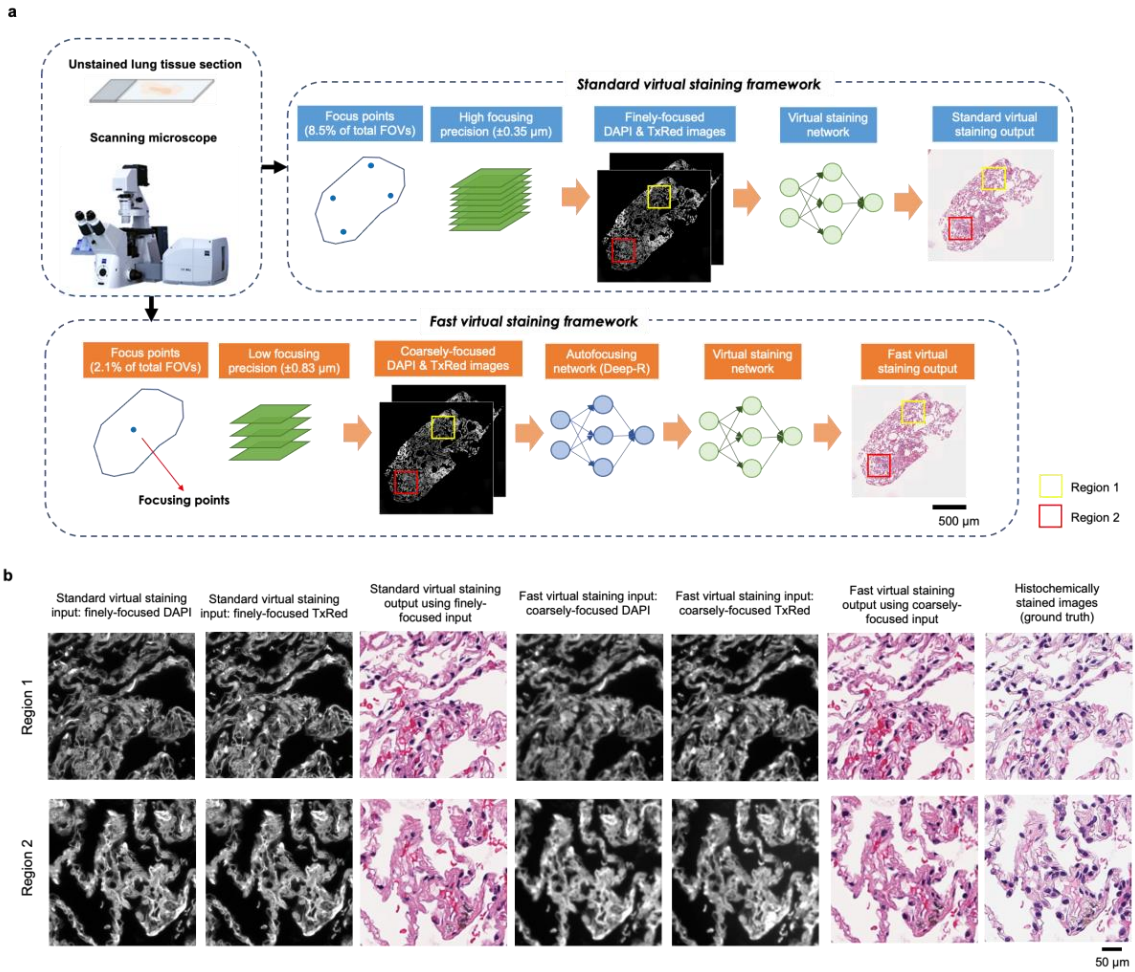

**Figure 2 Defocused image virtual staining**. **a**. Comparison of the standard in-focus image virtual staining framework and our defocused image virtual staining framework. The defocused image virtual staining framework can decrease 88.8% of the autofocusing time and 32.1% of total image acquisition time per WSI by using 4× fewer focus points and 2× lower focusing precision. **b**. Zoomed-in regions of the acquired autofluorescence images and the generated virtually stained images using the two frameworks in **a**. Histochemically stained ground truth images are also shown in the last column for comparison.

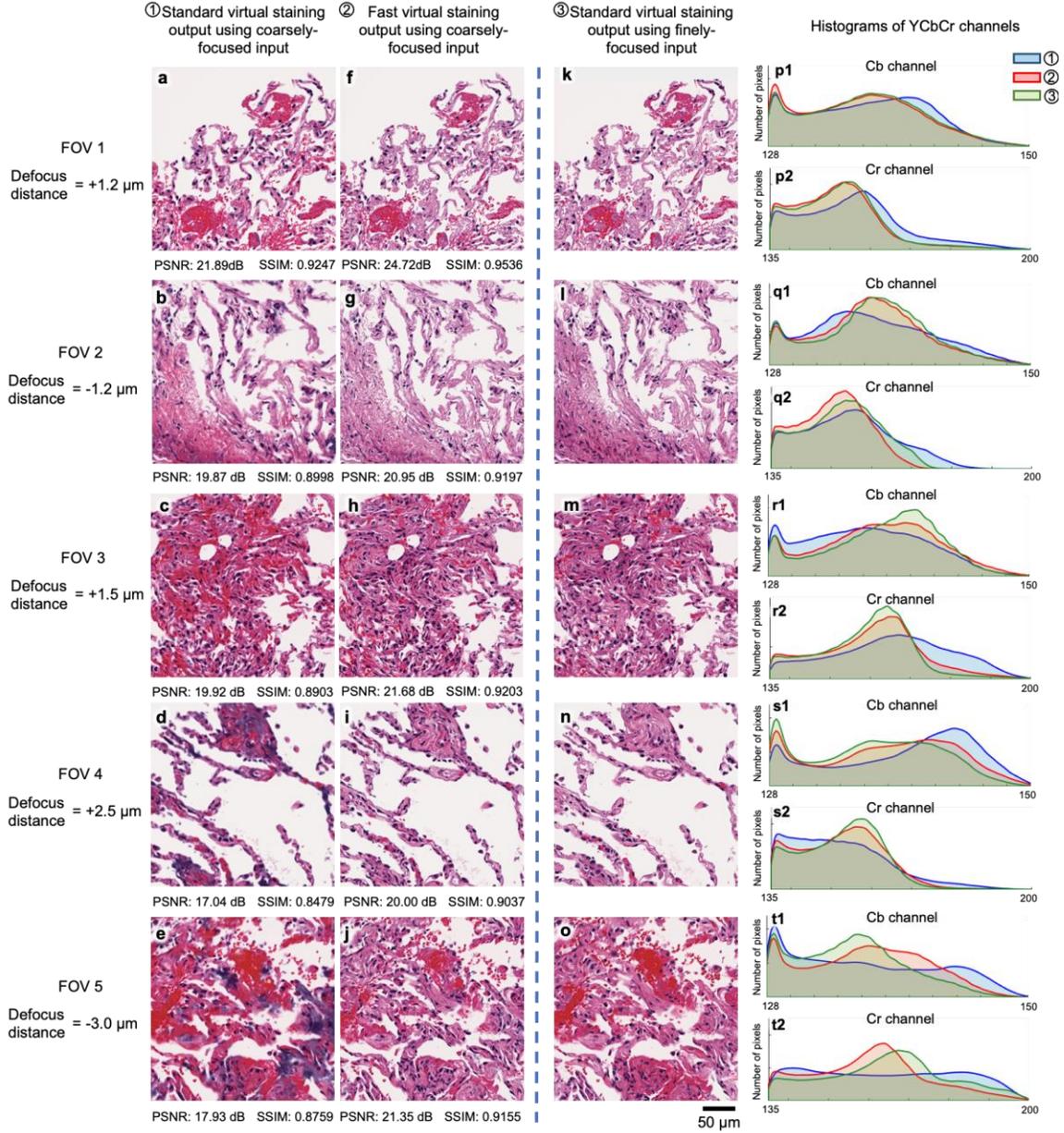

**Figure 3 Comparison between the results of the standard in-focus image virtual staining network and the defocused image virtual staining network using coarsely-focused input images**. **a-e.** Standard in-focus image virtual staining network output using coarsely-focused autofluorescence images as inputs on FOVs of different lung tissue sections that were never used during the training phase. **f-j.** Defocused image virtual staining output using the same coarsely-focused input images as **a-e**. **k-o.** Standard in-focus image virtual staining network output using finely-focused image FOVs are shown as baseline, for comparison purposes. **p-t.** Histograms of the Cb and Cr color channels for the virtual staining results in **a-e**, **f-j**, and **k-o**.

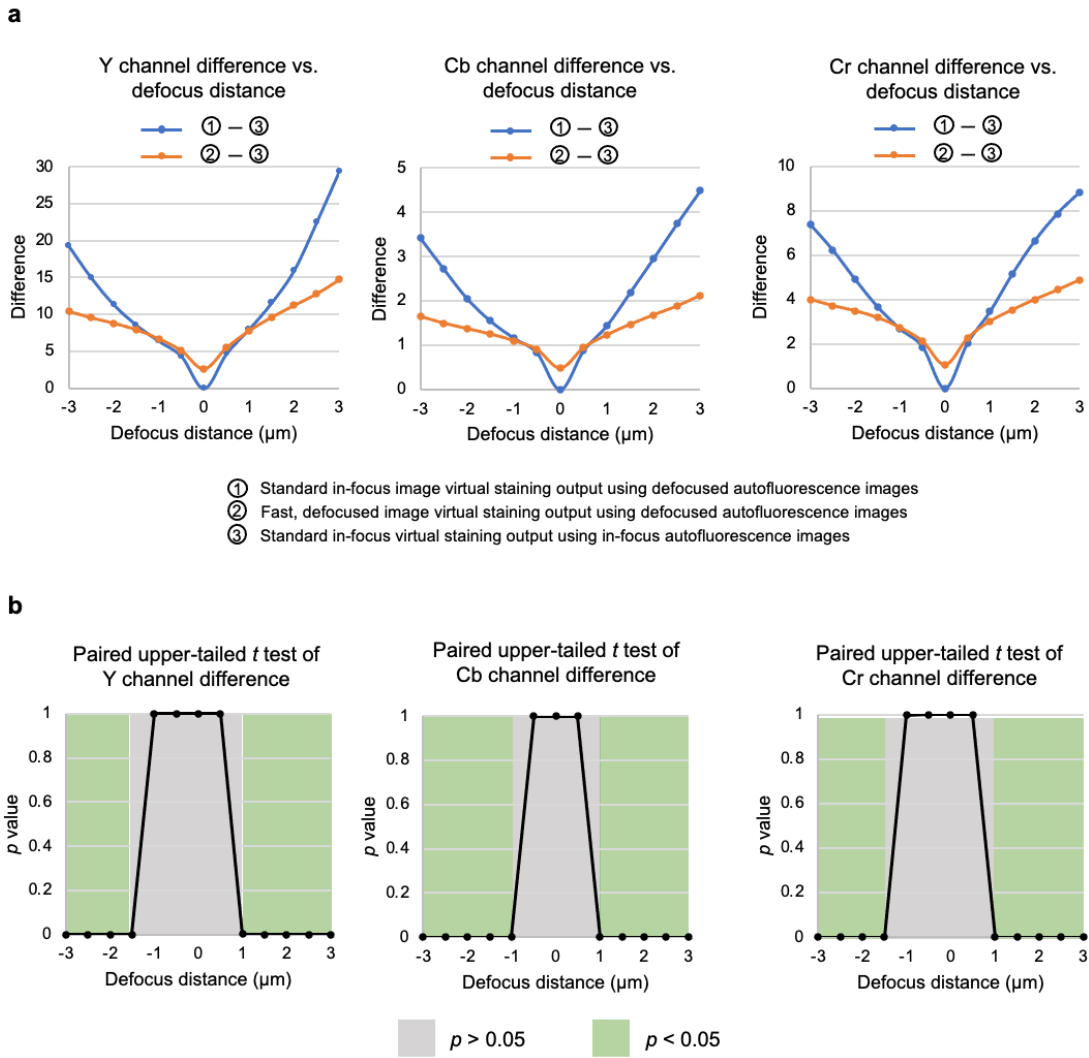

**Figure 4 Comparison of YCbCr color difference vs. the axial image defocus distance. a**. Absolute YCbCr color differences between (1) the standard in-focus image virtual staining (and the defocused image virtual staining) inference results on autofluorescence input images acquired at different axial depths and (2) the results of the standard virtual staining image inference using in-focus autofluorescence images are plotted as a function of the axial defocus distance. **b**. Results of the paired upper-tailed *t* tests (see the Materials and Methods section). The resulting *p* values are plotted as a function of the axial defocus distance. In the area of $p < 0.05$, the defocused image virtual staining framework statistically significantly improved virtual staining quality over the in-focus image virtual staining framework for the same defocused input autofluorescence images.

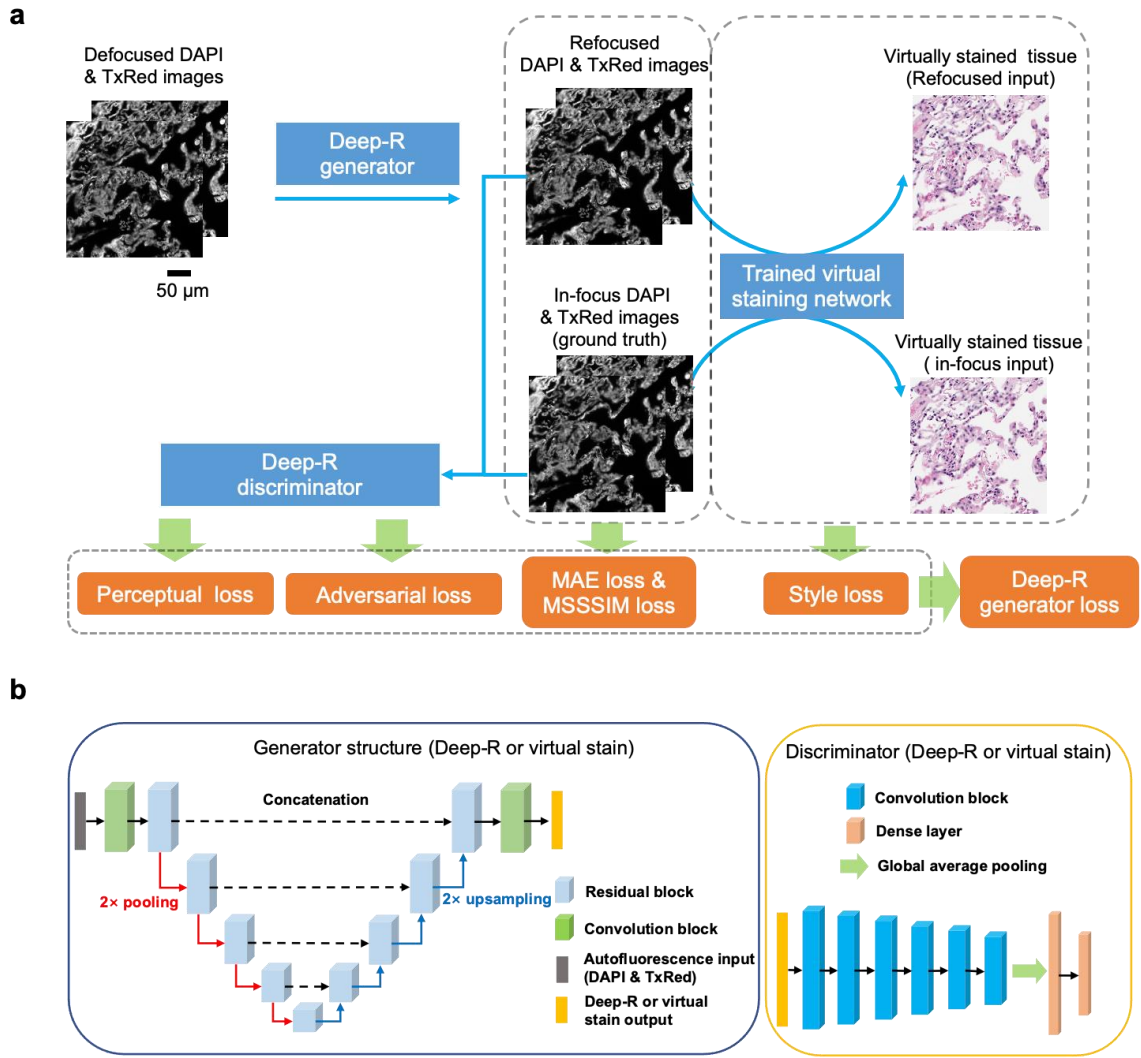

**Figure 5 Training loss and network architecture of the defocused image virtual staining framework**. A GAN architecture consisting of two deep neural networks including a generator and a discriminator was used to train the Deep-R and the virtual staining networks. **a**. Training details of the Deep-R network. The objective function of the Deep-R generator training contains a perceptual loss and an adversarial loss from discriminator, the MAE loss and the MSSSIM loss between Deep-R output and the ground truth in-focus images, and a style loss based on high-level image features of the trained virtual staining network (see the Materials and Methods section). **b**. The network architectures of the generator and discriminator used in the Deep-R and virtual staining networks.